\documentclass[sigconf]{acmart}

\copyrightyear{2026}
\acmYear{2026}
\setcopyright{cc}
\setcctype{by}
\acmConference[AISec '26]{19th Workshop on Artificial Intelligence and Security}{November 15--19, 2026}{The Hague, Netherlands}
\acmBooktitle{19th Workshop on Artificial Intelligence and Security (AISec '26), November 15--19, 2026, The Hague, Netherlands}
\acmDOI{10.1145/3847352.3848091}
\acmISBN{979-8-4007-3031-3/2026/11}

\usepackage[utf8]{inputenc}
\usepackage[T1]{fontenc}
\usepackage{hyperref}
\usepackage{xcolor}
\usepackage{booktabs}
\usepackage{tabularx}
\usepackage{enumitem}
\usepackage{graphicx}
\usepackage{amsmath}
\usepackage{natbib}
\usepackage{csquotes}
\usepackage{listings}
\usepackage{caption}
\usepackage{array}
\usepackage{tikz}
\usepackage{pifont}

\usetikzlibrary{arrows.meta,positioning,fit,backgrounds}

\definecolor{codegray}{rgb}{0.95,0.95,0.95}
\definecolor{codeblue}{rgb}{0.0,0.3,0.6}
\newcommand{\sys}{InferD}
\newcommand{\principle}{\emph{deception without inference}}
\newcommand{\cmark}{\ding{51}}
\newcommand{\xmark}{\ding{55}}

\newcommand{\obswindow}{Eight weeks (May-July 2026)}     %
\newcommand{\nreq}{24{,}192}           
\newcommand{\nsrc}{2{,}342}            
\newcommand{\ncountry}{61}
\newcommand{\nasn}{254}
\newcommand{\pctsingle}{71\%}
\newcommand{\pcttoptwo}{64\%}
\newcommand{\pcttopfiveasn}{61\%}
\newcommand{\ncoded}{601}              
\newcommand{\nbuilds}{2}               
\newcommand{\nsvc}{14}                 
\newcommand{\nlitepath}{14}            

\begin{document}

\title{InferD: A Honeypot Framework for AI Inference Infrastructure}

\author{Hicham Hammouchi}
\affiliation{%
  \institution{SnT, University of Luxembourg}
  \city{Luxembourg}
  \country{Luxembourg}
}
\email{hicham.hammouchi@uni.lu}

\author{Gabriele Lenzini}
\affiliation{%
  \institution{SnT, University of Luxembourg}
  \city{Luxembourg}
  \country{Luxembourg}
}
\email{gabriele.lenzini@uni.lu}

\renewcommand{\shortauthors}{Hammouchi and Lenzini}

\begin{abstract}
    Large language model serving stacks, commercial inference APIs, vector databases, and agent orchestration gateways are now exposed on the public internet and indexed by commodity scanners within hours of deployment. The threat characteristics of this surface are only beginning to be measured, and the honeypot frameworks that could measure it were designed for conventional network services rather than for AI infrastructure. In fact, emulating an AI service is a distinct deception problem: the observable behavior of the service, including model inventories, token formats, tool-call protocols, and authorization flows, must be reproduced faithfully, while the underlying computation, model inference and tool execution, must never occur. We present InferD, a low-interaction honeypot framework organized around a single principle we call \principle{}: every request an adversary can observe is answered from static response pools, and nothing an adversary submits is ever executed. \sys{} decomposes the emulation into four deception planes, transport, service protocol, vulnerability surface, and agentic session, adds a provenance plane that links a planted credential to evidence of downstream model-mediated processing.

\end{abstract}






\maketitle

\section{Introduction}
\label{sec:intro}
 
A honeypot is a network-accessible resource with no production purpose, placed where adversaries will find it so that any interaction becomes evidence. The value of the idea has always come from a good match between what the decoy imitates and what the community wants to learn: stack emulation for network reconnaissance, shell emulation for intrusion behavior, protocol emulation for industrial threats \citep{provos2004honeyd,spitzner2003honeypots}. The infrastructure that serves machine-learning inference is a surface the existing decoys were not built for, and it is now exposed at scale. Operators self-host model servers and notebook environments, run vector databases and experiment trackers, and stand up agent gateways that use protocols such as the Model Context Protocol (MCP) \footnote{Anthropic Model Context Protocol Specification: \url{https://modelcontextprotocol.io/specification}}; commercial inference is reached through a small set of well-known API shapes. Internet-wide scanners index these endpoints within hours \citep{durumeric2013zmap}, and the very features that identify an exposed inference service, including a version string, a model inventory, and an authorization header format, are the signals an adversary uses to build a target list.
 
Emulating this surface is not a matter of porting a honeypot to new ports. The behavior an adversary probes is the behavior of a model, and this creates a tension that sits at the center of the design. To be convincing, the facade must answer everything an adversary can observe: it must return a plausible model inventory, produce well-formed completions with the right streaming semantics, accept and validate credentials in the right formats, and carry an agentic session through its tool-call lifecycle. To be safe and useful as an instrument, it must run nothing an adversary submits, because the computation that would make the service real, executing a model, following a tool call, resolving a submitted URL, is the capability an attacker is trying to reach. A naive way to resolve the tension would be to place a real model behind the facade, as recent language-model-driven honeypots do for shells. For an \emph{inference} honeypot this is a deliberate design choice with significant tradeoffs: generation reintroduces serving cost and latency variance on the exact axis an adversary can measure, makes responses non-deterministic and hard to reproduce for analysis, and turns the honeypot into a live model that an adversary can prompt, jailbreak, and manipulate, which is precisely the artifact whose abuse the honeypot is meant to study from the outside. A model-backed facade may be appropriate for studying model-level attack behavior in other contexts, but for the goals of this work it is the wrong choice.
 
We resolve the tension with a principle we call \principle{}. Every observable behavior is served from static, pre-authored response pools, and the computation each request implies is logged as intent rather than carried out. This is the organizing idea of \sys{}, and the paper is largely an argument that it is the right idea and an account of what follows from taking it seriously. Two consequences do most of the work. First, a system that never needs to compute on attacker input has no legitimate reason to execute it or to reach the network on its behalf, which lets us pair non-execution by design with an enforceable operating-system property, a kernel egress rule that prevents any relay or exfiltration. Second, the same principle applies uniformly across a heterogeneous surface: a model-pull request, a chat completion, a vector upsert, and an MCP tool call are all answered from pools and all recorded as intent, so a single framework can span self-hosted services and cloud API facades without special-casing any of them.
 
Our contribution is a design and the reasoning behind it. We state the \principle{} principle and develop its consequences as a layered architecture, treating each deception mechanism as a problem of faithful emulation weighed against the alternative it displaces and the way an adversary could detect it. We show how the principle yields two complementary safety guarantees, non-execution of submitted input by construction and kernel-enforced egress containment that holds even if an emulator misbehaves, and we document the systems techniques, source attribution under container networking, egress isolation bound to the emulator's operating-system user, and response-pool construction, that make the design safe and faithful on commodity hardware. We ground the emulated surface in the record of attributable, exploited-in-the-wild vulnerabilities so that its scope is principled and reproducible, and we make its coverage of the agentic surface legible by mapping it onto an external risk taxonomy. Finally, we report an initial deployment that validates the pipeline end to end and surfaces a first empirical finding: a purpose-built cloud-credential harvester delivered via an MCP stdio code-execution primitive, observed from three sources and captured with zero code executed.
 
The paper is organized as follows. Section~\ref{sec:problem} sets out why the surface exists and is exposed, what makes its emulation distinct, and the threat model and design goals that follow. Section~\ref{sec:principle} develops the design: the \principle{} principle, the architecture, and the deception planes. Section~\ref{sec:impl} documents the implementation and safety guarantees. Section~\ref{sec:deploy} reports the deployment, including a case study of a captured credential-harvesting payload. Section~\ref{sec:discussion} covers detectability limits, extensibility, and related work. Section~\ref{sec:conclusion} concludes.

\section{Background and Threat Model}
\label{sec:problem}
 
\subsection{Structural Origins of Exposure}
 
That so much inference infrastructure is reachable from the open internet is not simply carelessness but rather a predictable consequence of how the software distributes its own threat model. A recurring pattern in AI serving frameworks is to ship with a security posture that presumes a trusted network and to delegate the residual isolation requirement to the operator. Concretely: Ollama \footnote{Ollama: Run Large Language Models Locally. \url{https://ollama.ai}} binds to loopback by default, but its documented multi-user configuration sets \texttt{OLLAMA\_HOST=0.0.0.0}, making network-wide exposure the recommended deployment posture; Ray's cluster dashboard is documented as unauthenticated by design, with the explicit assumption that a cluster-level firewall supplies the missing boundary. When the operator does not supply that isolation, the result is not an aberration but the predictable default outcome. Internet-wide measurement confirms this is not anecdotal: a study of 320,102 publicly reachable LLM services across 15 deployment frameworks found widespread absence of authentication and TLS, with unauthenticated endpoints concentrated on the same well-known ports that scanners trivially enumerate \citep{hou2025landscape}.\footnote{Industry scanning of approximately one million exposed AI endpoints corroborates these findings at larger scale; see \url{https://thebrightbyte.com/playbook/insights/one-million-exposed-ai-services-2026}.} Ray's cluster API was exploited at this scale in the ShadowRay campaign\footnote{Oligo Security, \textit{ShadowRay: First Known Attack Campaign Targeting AI Workloads Actively Exploited in the Wild (CVE-2023-48022)}, 2024. \url{https://www.oligo.security/blog/shadowray-attack-ai-workloads-actively-exploited-in-the-wild}}, and the Probllama path-traversal flaw became remotely exploitable with root privileges precisely because Ollama's recommended configuration binds to all interfaces\footnote{Wiz Research, \textit{Probllama: Ollama Remote Code Execution Vulnerability (CVE-2024-37032)}, 2024. \url{https://www.wiz.io/blog/probllama-ollama-vulnerability-cve-2024-37032}}. The decoy needs only to present the observable fingerprint of a default, network-exposed deployment, a posture the broader ecosystem produces routinely, whether through deliberate exposure or absent isolation.
 
\subsection{Emulation Constraints Specific to AI Services}
 
Three properties separate this surface from the services classical honeypots imitate. First, fidelity is defined against the behavior of a model. A model server does not merely return a banner; it advertises an inventory, negotiates streaming, and produces content whose shape an adversary can inspect. Second, part of the surface is stateful and agentic. A single request from an automated agent reveals little, but a maintained session of tool calls exposes structured intent, so an instrument that captures only isolated requests misses the most informative behavior. Third, part of the surface is reached through credentials rather than through open ports, which shifts the observation problem from \enquote{who connects} to \enquote{who possesses a key and what they do with it}. Each property pulls toward higher interaction, and each therefore sharpens the tension between fidelity and safety that the next section resolves.
 
\subsection{Threat Model and Design Goals}
 
We design \sys{} against five adversary classes, ordered by increasing sophistication and decreasing prevalence. \textbf{A1, indiscriminate scanners} issue a small fixed set of probes to every reachable host and rarely return. \textbf{A2, fingerprinters} enumerate inventories, versions, and token formats to catalog exposed capacity. \textbf{A3, credential abusers} present keys, stuffed or recovered, against inference facades to obtain unauthorized model access. \textbf{A4, exploit operators} send payloads crafted for specific software flaws. \textbf{A5, agentic actors} maintain multi-step sessions, issue tool calls, or route captured output back through a model, exposing intent that a single request does not.
 
The design is bounded first by what the instrument must never do, treated as constraints rather than options: it must not execute submitted code, forward or evaluate tool calls, run a model, emit outbound traffic, or disclose the real addresses of the sources it observes. From these constraints and the adversary classes we derive six design goals, named and referenced throughout. \textbf{G1, protocol fidelity}: requests and responses should be indistinguishable from the real service. \textbf{G2, non-execution}: the system records but never performs the computation an attacker attempts to induce. \textbf{G3, source attribution}: the true origin of each request is captured despite proxying and container networking; this is non-trivial because standard container port mapping silently rewrites every source address to the container gateway IP, producing a dataset that appears complete but records the wrong address for every event. \textbf{G4, stateful capture}: agentic interactions persist across requests. \textbf{G5, provenance}: a planted credential can be linked to its first use and, where possible, to downstream model-mediated processing. \textbf{G6, fingerprint resistance}: responses carry enough controlled diversity that the facade is not trivially cataloged from repetition.

\section{Design}
\label{sec:principle}
 
The design has three parts. The \principle{} principle determines what gets answered and what gets withheld: a static response for every observable request, and a log entry for every implied computation. The architecture determines where emulators sit and how traffic reaches them: a two-surface model separating open-port services from credential-gated cloud facades, unified under a single egress boundary. The deception planes determine how each observable layer of an AI service is reproduced, with each plane developed against the alternative it rejects and the detection vector it accepts. The three parts are described in the subsections that follow.
 
\subsection{The Deception-Without-Inference Principle}
 
The principle states that faithful protocol behavior can be delivered entirely from static response pools, while the computation each request implies is recorded and withheld.
 
The alternative to static pools is generation: place a model, or a fine-tuned smaller model, behind the facade and let it answer, as language-model-driven shell honeypots do to obtain unbounded interaction depth \citep{sladic2024llm,otal2024llm}. For a shell this is attractive because depth is the point. For an inference honeypot, however, generation reintroduces serving cost and latency variance on the exact axis an adversary can measure; it makes responses non-deterministic and therefore hard to reason about or to reproduce for analysis; and, most importantly, it turns the honeypot into a live model that an adversary can prompt, jailbreak, and manipulate, which is the artifact whose abuse the honeypot is meant to study from the outside. Static pools invert every one of these properties. They are cheap and constant in latency, they are deterministic and auditable, and they present no model to attack. Besides, retrieval-augmented designs such as DecoyPot \citep{sezgin2025decoypot} narrow the hallucination gap but still infer at request time, so the measurable latency and the non-execution guarantee remain unmet. The cost they pay is finite response depth: a pool cannot answer an adversarial prompt whose correct reply depends on genuine reasoning. We accept this cost knowingly, mitigate its most visible symptom (repetition) with variant expansion (Section~\ref{sec:pools}), and note that it is detectable only by an adversary who specifically tests for reasoning, a small and identifiable minority of the traffic that reaches such a surface. We regard the finite depth not as a weakness of the instrument but as the price of a much stronger position, which the principle buys almost for free: a facade that never computes on attacker input has no legitimate reason to reach the network, so the design's non-execution can be backed by a kernel-enforced egress guarantee that no emulator bug can void (Section~\ref{sec:safety}).
 
The principle also gives the framework its uniformity. Because every emulated service is an instance of the same idea, answer from a pool and record the implied computation, a single middleware and storage path can serve a heterogeneous surface without per-service trust decisions, which is what makes it sensible to speak of a framework rather than a collection of decoys.
 
\subsection{Architecture and the Two-Surface Model}
 
\begin{figure*}[t]
\centering
\begin{tikzpicture}[
  node distance=0mm,
  font=\scriptsize,
  arr/.style={-{Stealth[length=2mm,width=1.4mm]}, line width=0.7pt, gray!55!black},
  inet/.style={
    draw=gray!45, fill=gray!12, rounded corners=3pt,
    minimum width=114mm, minimum height=7mm, align=center
  },
  surf/.style={
    draw=#1!45, fill=#1!8, rounded corners=3pt,
    minimum width=54mm, minimum height=15mm,
    align=center, inner sep=3pt, text width=50mm
  },
  mcomp/.style={
    draw=#1!55!black, fill=#1!7, rounded corners=3pt,
    minimum width=114mm, minimum height=8mm,
    align=center, inner sep=2pt, text width=110mm
  },
]

\node[inet] (adv)
  {\bfseries Internet probe
   \hspace{3mm}\normalfont\itshape adversary classes A1--A5};

\node[surf=blue, anchor=north]
  (s2) at ([xshift=-30mm, yshift=-5mm]adv.south) {%
  {\color{blue!60!black}\bfseries Surface 2}\ {\color{black!45}\scriptsize TLS :443}\\[2pt]
  \color{black!65}cloud API facades\\[1pt]
  \color{black!50}\footnotesize OpenAI - Anthropic - Gemini\\
  \footnotesize Assistants - MCP};

\node[surf=teal, anchor=north]
  (s1) at ([xshift=30mm, yshift=-5mm]adv.south) {%
  {\color{teal!55!black}\bfseries Surface 1}\ {\color{black!45}\scriptsize direct ports}\\[2pt]
  \color{black!65}self-hosted services\\[1pt]
  \color{black!50}\footnotesize Ollama - vLLM - LiteLLM\\
  \footnotesize Qdrant - Jupyter - MLflow -  Gradio - OpenClaw};

\node[mcomp=orange, anchor=north]
  (mw) at ([xshift=30mm, yshift=-5mm]s2.south) {%
  {\color{orange!60!red!75}\bfseries Middleware}\\[1pt]
  \color{black!70}source attribution (G3)\quad - \quad
  structural anomaly scan\quad - \quad canary-echo scan (G5)};

\node[mcomp=green, anchor=north]
  (emu) at ([yshift=-1.5mm]mw.south) {%
  {\color{green!35!black}\bfseries Per-service emulator}\\[1pt]
  \color{black!70}auth check\quad - \quad
  static response pool (G1, G6)\quad - \quad
  honeytoken detection\quad - \quad canary injection};

\node[draw=gray!50, fill=gray!13, rounded corners=3pt,
      minimum width=114mm, minimum height=7mm,
      align=center, inner sep=2pt, text width=110mm,
      anchor=north]
  (log) at ([yshift=-1.5mm]emu.south) {%
  {\color{gray!50!black}\bfseries Canonical event log}};

\begin{scope}[on background layer]
\node[draw=red!50!black, dashed, rounded corners=4pt, line width=0.6pt,
      fit=(mw)(emu)(log), inner sep=4pt,
      label={[font=\tiny\itshape, text=red!50!black, yshift=-1mm]above right:%
             egress-isolated: emulator user,\ kernel \texttt{OUTPUT DROP}\ (G2)}] {};
\end{scope}

\draw[gray!55!black, line width=0.7pt]
  (adv.south) -- ++(0,-2.5mm) coordinate (vsplit);
\draw[arr] (vsplit) -| (s2.north);
\draw[arr] (vsplit) -| (s1.north);

\coordinate (smconv) at ([yshift=2.5mm]mw.north);
\draw[gray!55!black, line width=0.7pt]
  (s2.south) -- ++(0,-1.5mm) -| (smconv);
\draw[gray!55!black, line width=0.7pt]
  (s1.south) -- ++(0,-1.5mm) -| (smconv);
\draw[arr] (smconv) -- (mw.north);

\draw[arr] (mw.south) -- (emu.north);
\draw[arr] (emu.south) -- (log.north);

\end{tikzpicture}
\caption{\sys{} capture path. All requests traverse common middleware and a
per-service emulator before reaching an append-only log, inside a kernel-level
egress boundary}
\Description{Architecture of InferD. Internet probes reach either
Surface 1, containing directly exposed self-hosted AI services, or
Surface 2, containing TLS-fronted cloud API facades. Both paths pass
through common middleware and a per-service emulator before being
recorded in the canonical event log. The complete capture path is
protected by a kernel-enforced outbound-traffic boundary.}
\label{fig:arch}
\end{figure*}
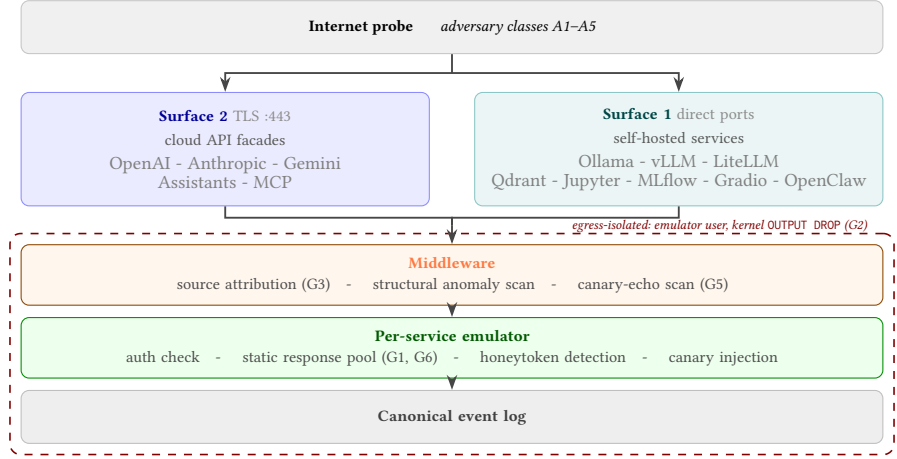

\sys{} runs as a single host-networked container on a commodity server. A reverse proxy terminates TLS for the cloud API facades and forwards plaintext to an internal port; the self-hosted services bind directly on their conventional ports. All traffic passes through common middleware before reaching a per-service emulator, and the whole application runs inside a kernel-level egress boundary (Figure~\ref{fig:arch}).
 
A design decision that recurs through the paper is to treat the surface as two structurally different parts rather than one, because they attract different adversaries and demand different instrumentation. \emph{Surface~1} comprises self-hosted services discovered through open ports; the deception target is the default, network-exposed deployment described in Section~\ref{sec:problem}, and the capture of interest is reconnaissance, inventory enumeration, and vulnerability probing. \emph{Surface~2} comprises cloud API facades reached not by scanning but by an adversary who already holds a credential; the deception target is a genuine software-as-a-service endpoint, and the capture of interest is credential abuse and the content of the prompts and tool definitions an authenticated caller submits. Keeping the two apart lets the provenance plane attach specifically to Surface~2, where credentials are the entry mechanism, without complicating the open-port emulators where they are not. Table~\ref{tab:surface} lists the emulated facades; each is an instance of \principle{}.
 
\begin{table*}[h]
\centering
\footnotesize
\caption{Emulated service facades across the two surfaces.}
\label{tab:surface}
\renewcommand{\arraystretch}{1.2}
\begin{tabularx}{\textwidth}{@{}l l X@{}}
\toprule
\textbf{Service} & \textbf{Port / path} & \textbf{Primary capture} \\
\midrule
\multicolumn{3}{@{}l}{\textit{Surface 1: self-hosted, open ports}}\\
Ollama            & 11434 & Model enumeration and pull, generation probes \\
vLLM              & 8001  & Completions deserialization and multimodal RCE payloads \\
HuggingFace TGI   & 8080  & Token-format probes, model enumeration \\
LiteLLM           & 4000  & MCP stdio test-connection probes, credential-harvesting payloads, SSRF (CVE-2024-6587) \\
Qdrant            & 6333  & Collection enumeration, poisoning writes, exfiltration \\
Jupyter           & 8888  & Token brute force, path enumeration, code payloads \\
MLflow            & 5000  & Credential stuffing, experiment and artifact probes \\
Gradio            & 7860  & Prompt injection via the predict endpoint \\
OpenClaw          & 18789 & Localhost-trust bypass, privilege escalation, unauthenticated admin, skill-package URL probes \\
\midrule
\multicolumn{3}{@{}l}{\textit{Surface 2: cloud API facades, TLS on :443}}\\
OpenAI-compatible & \texttt{/v1/}          & Key stuffing, injection, enumeration \\
Anthropic         & \texttt{/v1/messages}  & Key-format probes, multi-turn payloads \\
Gemini            & \texttt{/v1beta/models/}& Key probes, safety-bypass attempts \\
Assistants        & \texttt{/v1/threads}   & Tool definitions, run reasoning \\
MCP               & \texttt{/mcp}          & Tool-call sequences, OAuth flows \\
\bottomrule
\end{tabularx}
\end{table*}
 
\subsection{Deception by Layer}
 
The principle becomes a system through a stack of independent mechanisms, each operating at a different level of what the adversary can observe. We develop each as a problem of faithful emulation, state the alternative we considered and rejected, and name the way an adversary could detect the deception. The recurring shape of these limits, that non-execution is detectable to anyone who tests for effect, and is a trade-off of the design.
 
\subsubsection{Transport Layer}
 
The problem is to record the true origin of each request despite the reverse proxy in front of Surface~2, and to present transport characteristics consistent with the imitated service. Running the container in host-network mode preserves the genuine peer address on the direct ports, and for proxied traffic the middleware trusts the \texttt{X-Forwarded-For} header only from the loopback proxy peer, ignoring it from any other source; the mechanism is detailed in Section~\ref{sec:networking}. The alternative, standard container port mapping, is what a straightforward deployment would use, and it silently rewrites every source address to the container gateway IP, with no error or warning, so the dataset appears complete but records the same gateway address for every event; we treat avoiding this as a correctness requirement, not an optimization. The residual limit is that a determined adversary can compare the transport fingerprint against the genuine provider and may find discrepancies; we log the fingerprint so that any such mismatch is visible in our own data as well.
 
\subsubsection{Service Protocol Layer}
 
Here the adversary reads what the service advertises and the shape of what it returns, and fidelity (G1) is the reproduction of that behavior without computing it. \sys{} serves each endpoint from a static pool and diversifies a small number of fields per response through variant expansion (Section~\ref{sec:pools}), so that repeated requests do not return byte-identical replies. The residual limit is the finite depth of a pool: an adversary who submits a prompt with a checkable answer will find that the facade does not reason. This limit is real and we do not paper over it, but it binds only the most capable adversaries and is, usefully, itself a detector of them.
 
\subsubsection{Vulnerability Surface Layer}
 
Exploit operators (A4) send payloads crafted for specific flaws, and to capture those payloads the facade must present the vulnerable-looking interface while the flaw itself does not exist. The emulated surfaces combine two categories: interfaces grounded in attributable CVEs from the incident record, which ensures the bait matches payloads observed in the wild, and structurally motivated surfaces for services whose architectural properties make specific attack patterns likely even without a single documented incident. The set is not exhaustive; the pool-and-emulator architecture is designed to accommodate additional surfaces as the CVE record and threat landscape evolve. Current CVE-grounded surfaces include: a model-pull endpoint that accepts an attacker-supplied registry path, mirroring the Ollama path traversal of CVE-2024-37032 (documented above); a completions field that a vulnerable vLLM build would deserialize when processing multimodal inputs; a gateway request that a vulnerable build would forward with its upstream credential attached, mirroring the server-side request forgery \citep{pellegrino2016uses} of CVE-2024-6587\footnote{National Vulnerability Database, \textit{CVE-2024-6587: Server-Side Request Forgery in LiteLLM}, 2024. \url{https://nvd.nist.gov/vuln/detail/CVE-2024-6587}}; and an MCP \texttt{test/connection} endpoint that a vulnerable LiteLLM \footnote{LiteLLM: Call 100+ LLM APIs using the OpenAI format. \url{https://github.com/BerriAI/litellm}} build would use to spawn an attacker-specified subprocess. Alongside these signature-shaped surfaces the middleware runs a structural anomaly scan, independent of any single vulnerability, examining each request for shell metacharacters, embedded code, unexpected URL fields, and path traversal, so that payloads whose significance is not yet known are preserved for retrospective analysis. The structural anomaly scan proved its value beyond signature-matched CVEs in the deployment: it correctly identified a novel credential-harvesting payload delivered via the LiteLLM MCP test-connection feature, a pattern not matching any single known CVE at capture time, and preserved the full payload for the retrospective analysis in Section~\ref{sec:payload}. 
 
\subsubsection{Agentic Session Layer}
 
A single request from an agentic actor (A5) reveals little; a maintained session of tool calls reveals structured intent, so this layer is where statefulness (G4) earns its cost. \sys{} carries two stateful surfaces: an assistants interface \footnote{OpenAI Assistants API Documentation: \url{https://platform.openai.com/docs/assistants}} whose threads expose a create-run-submit lifecycle, and an MCP surface that implements the handshake, authorization discovery, dynamic client registration, and a token endpoint, holding session state so that a tool-call sequence and its arguments are recorded together. Tool calls are logged as arguments and never dispatched, which keeps the layer inside the non-execution guarantee. The agentic capture machinery is designed and implemented; its empirical validation against organic A5 traffic is limited in the current window, as discussed in Section~\ref{sec:deploy}. We frame the intent these sessions expose using the lethal trifecta, Willison's formulation\footnote{Simon Willison, \textit{The Lethal Trifecta for AI Agents} (2025). \url{https://simonwillison.net/2025/Jun/16/the-lethal-trifecta/}} under which an agentic attack requires the co-occurrence of untrusted instructions, access to sensitive data, and an exfiltration path. The honeypot instruments the untrusted-instruction and exfiltration-path legs while structurally withholding the data leg. This framing is more than descriptive. Systematic evaluation shows that current agents comply with instructions embedded in tool descriptions at substantial rates, that susceptibility rises with model capability, and that safety alignment offers little protection before execution \citep{wang2025mcptox,gasmi2025mcp}, which is the reason to record tool-call arguments verbatim rather than to rely on the agent to refuse.
 
Because coverage of an emerging surface is easy to assert and hard to substantiate, we state it against an external taxonomy, the OWASP MCP Top 10 \footnote{\url{https://owasp.org/www-project-mcp-top-10/}}, a community project to catalog the most critical security risks in MCP tool integration. We use this MCP-specific taxonomy rather than the broader OWASP Top 10 for LLM Applications\footnote{\url{https://owasp.org/www-project-top-10-for-large-language-model-applications/}} because the agentic session layer's capture is defined at the protocol level MCP introduces, not at the level of general LLM application risk. Table~\ref{tab:owasp} maps \sys{}'s coverage onto all ten OWASP MCP Top 10 categories, reporting whether each is captured (\cmark{}), outside the passive inbound observation scope (\xmark{}), or partially observable (\textasciitilde). The mapping is pointed in two places: category MCP05 (command injection and execution) is now directly demonstrated rather than claimed, as the LiteLLM credential-harvesting payload in Section~\ref{sec:payload} is precisely an MCP05 exploitation attempt captured in the wild; and category MCP08 (lack of audit and telemetry) names a passive honeypot as a source of precisely the tool-invocation telemetry whose absence it describes, so the instrument's function coincides with a named gap in the standard.

\begin{table*}[h]
\centering
\footnotesize
\caption{\sys{} coverage of the OWASP MCP Top 10 (beta, MCP01:2025--MCP10:2025). \cmark{} observed in inbound traffic; \xmark{} outside passive scope; \textasciitilde{} partially observable; \dag{} covered as a capability, not an observed attack.}
\label{tab:owasp}
\renewcommand{\arraystretch}{1.25}
\setlength{\tabcolsep}{5pt}
\begin{tabular}{@{}p{0.23\textwidth} p{0.61\textwidth} >{\centering\arraybackslash}p{0.09\textwidth}@{}}
\toprule
\textbf{OWASP MCP category} & \textbf{\sys{} observation} & \textbf{Coverage} \\
\midrule
MCP01 Token mismanagement \& secret exposure & Honeytoken and credential capture; canary provenance & \cmark{} \\
MCP02 Privilege escalation via scope creep & OAuth scope and dynamic-registration requests & \cmark{} \\
MCP03 Tool poisoning & Poisoned tool descriptions and injection payloads & \cmark{} \\
MCP04 Software supply chain attacks \& dependency tampering & Supply-chain compromise occurs before network contact; not observable by passive inbound capture & \xmark{} \\
MCP05 Command injection \& execution & Command execution payloads captured via structural anomaly scan; demonstrated by the LiteLLM case study (Section~\ref{sec:payload}) & \cmark{} \\
MCP06 Intent flow subversion & Full tool-call sequences and arguments per session & \cmark{} \\
MCP07 Insufficient authentication \& authorization & Auth headers and token-endpoint interactions & \cmark{} \\
MCP08 Lack of audit and telemetry & Not an attack. \sys{} covers this gap by providing the verbatim tool-invocation telemetry the category names & \dag{} \\
MCP09 Shadow MCP servers & Detection requires active network discovery within a target environment; not visible from inbound traffic & \xmark{} \\
MCP10 Context injection \& over-sharing & Tool-call body content captured for injection analysis; cross-session context propagation not observable passively & \textasciitilde{} \\
\bottomrule
\end{tabular}
\vspace{2pt}\par\noindent\footnotesize\dag{} MCP08 is a defensive gap, not an attacker behavior. \sys{} covers it by supplying the missing telemetry; the \cmark{} rows denote attacks observed in captured traffic.\normalsize
\end{table*}
 
\subsubsection{Provenance Plane}
 
Observing that a credential was used is informative; tracing a specific planted credential from theft to first use, and further to evidence that captured output was processed by a model, is more so, and this is the provenance goal (G5). \sys{} seeds honeytokens, credentials whose only purpose is to be stolen \citep{spitzner2003honeytokens, shabtai2016behavioral}, and records on every request whether the presented credential matches a planted one and which token it is \cite{srinivasa2020towards}. A fraction of honeytoken responses embed a canary \footnote{Thinkst Canarytokens: \url{https://canarytokens.org}}, a unique marker with no legitimate reason to reappear; a later re-presentation of that marker is evidence that the captured output was ingested and re-emitted by a model-mediated pipeline rather than replayed. The plane's honest dependency, which the deployment made concrete, is that it produces signal only when credentials are actually found and used; a decoy that waits passively will observe credential abuse only once its keys are in circulation, a point we return to in Section~\ref{sec:deploy}.
 
\section{Implementation}
\label{sec:impl}
 
We describe the implementation with emphasis on the few places where a straightforward build would quietly do the wrong thing, since those are where the design's guarantees are actually won.
 
\subsection{Process Architecture}
\label{sec:procarch}
 
The facades are implemented as routers organized into ten process groups by port and transport, one group serving the TLS-fronted cloud facades and one each for the self-hosted services, each a separate application server managed by supervisord inside a single container. The arrangement is more than operational tidiness. Every group runs as the same unprivileged operating-system user, and that user is the one to which the egress rule of Section ~\ref{sec:egress-isolation} is bound, so a single kernel-level control covers all services regardless of which one an adversary reaches. Tying the process model to the safety boundary in this way means that adding a service cannot silently widen the trust surface: a new group inherits the same user and therefore the same egress prohibition.
 
\begin{lstlisting}[caption={Process definition (Ollama group). Every group runs as the egress-bound user.}]
[program:ollama]
command=uvicorn honeypot.services.ollama:app --host 0.0.0.0 --port 11434
user=honeypot        ; the OS user bound by the egress rule below
autostart=true
autorestart=true
\end{lstlisting}
 
\subsection{Source Attribution Under Container Networking}
\label{sec:networking}
 
Correct attribution (G3) is the most easily lost property in a containerized deployment. Standard port mapping routes traffic through a userland proxy, so the application sees every connection as originating from the container gateway and the true address is rewritten before it arrives; the failure is silent because no error is raised and the application continues to log normally, with every recorded source address resolving to the same Docker bridge IP. The resulting dataset appears complete but is analytically worthless for any attribution, geographic, or campaign-correlation analysis. \sys{} runs in host-network mode so that each process observes the raw peer address at the host interface. For TLS connections the socket peer is always the loopback proxy, so the middleware reads the \texttt{X-Forwarded-For} header set by the reverse proxy in that case, and only from the loopback peer, which prevents an external caller from spoofing its own origin.
 
\subsection{Egress Isolation}
\label{sec:egress-isolation}
 
Container runtimes manipulate the host firewall directly, and higher-level firewall configuration does not intercept traffic the runtime has already permitted, so a container has unrestricted outbound access by default. For a honeypot that receives tool-call arguments containing attacker-specified URLs, this is the difference between an instrument and a relay. The mitigation binds the firewall to the operating-system user that runs every emulator (G2): loopback outbound is allowed so the reverse-proxy flow works, established connections are allowed so responses reach clients, and every other new outbound connection from that user is dropped, for IPv4 and IPv6, persisted across reboots:
 
\begin{lstlisting}[language=bash,caption={Kernel egress rules binding non-execution to the emulator OS user.}]
iptables  -A OUTPUT -m owner --uid-owner honeypot \
          -m state --state NEW ! -d 127.0.0.1 -j DROP
ip6tables -A OUTPUT -m owner --uid-owner honeypot \
          -m state --state NEW ! -d ::1 -j DROP
\end{lstlisting}
 
Because the boundary is enforced by the kernel, the relay-prevention property does not depend on the correctness of any emulator; a bug that attempted to follow an attacker URL could not leave the host. This is a stronger guarantee than a high-interaction honeypot can offer, since the latter attains realism by running the very software whose exploitation is documented in the incident record.
 
\subsection{Response Pools}
\label{sec:pools}
 
A response pool is a set of pre-authored objects for an endpoint; at response time the emulator draws one and expands a small number of marked fields so that identical requests do not yield identical replies, which is the concrete mechanism behind fingerprint resistance (G6) and, at the protocol layer, behind fidelity (G1). Pool size varies by endpoint: completion endpoints carry 40 to 60 variants, model-inventory endpoints carry 15 to 20 variants, and embedding endpoints are served from pools of unit-norm vectors at the standard dimensionalities so that an embeddings request returns vectors of the expected shape without any model. Variant expansion randomizes token counts, latency headers, request-id fields, and a small number of content tokens across each draw. A source sending more requests than the pool size will begin to observe content repetition; variant expansion defers byte-identical repetition by randomizing several fields per draw, but does not prevent a statistical fingerprint from emerging with sufficient samples, which we acknowledge as the concrete detectability limit discussed in Section~\ref{sec:discuss}.
 
\subsection{Non-Execution Guarantees and Ethical Constraints}
\label{sec:safety}
 
\sys{} makes two guarantees. Non-execution is a property of the design: every request is answered from a static pool, and no emulator spawns a process, evaluates code, loads a model, or dispatches a tool call, so payloads are logged as intent and never run. The decoy also holds no real model, key, or datastore, so nothing of value exists to compromise. As a backstop, the kernel egress rule of Section~\ref{sec:egress-isolation} drops all outbound connections from the honeypot user, so even a buggy emulator could not relay or exfiltrate.
 
This structural property also settles the ethical position. The prevailing view holds that passive defensive deception is permissible for security research when the operator does not initiate attacks against third parties, does not entrap those who would not otherwise act, and does not operate the honeypot as a relay \citep{sokol2017privacy,ngoma2024privacy}; \sys{} satisfies all three structurally rather than by policy, since it accepts only inbound connections, scripts every response, and blocks outbound traffic at the kernel. This design aligns with the ethical guidelines for passive internet measurement that Partridge and Allman \citep{partridge2016ethics} propose: the collection methodology does not require human-subject consent, does not interfere with third-party systems, and does not produce data obtainable by other means. Because the deception imposes no cost on anyone not already engaged in unauthorized access or automated scanning, the proportionality condition for deception-based research is met \citep{fraunholz2018demystifying}.
 
\section{Deployment and Findings}
\label{sec:deploy}
 
We deployed \sys{} on a single commodity server and operated it continuously for \obswindow{}. We present the deployment as a pipeline validation, not a measurement study, and we are deliberately conservative in what we read from it: a single-vantage observation over a short window can establish that the instrument functions and can suggest the shape of the traffic, but it cannot support population-level claims, and we do not make any.
 
\begin{figure}[h]
\centering
\includegraphics[width=1.\linewidth]{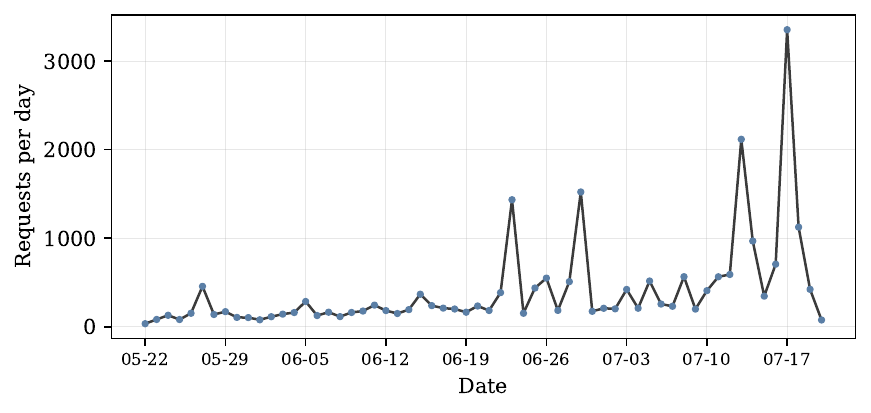}
\caption{Daily request volume over the deployment window}
\Description{Line chart showing the number of requests received per day during the eight-week deployment from May to July 2026. Traffic shows a generally low, slowly rising baseline interrupted by several short bursts of substantially higher request volume.}
\label{fig:daily}
\end{figure}

Two things the deployment establishes are worth stating plainly. First, the instrument works end to end: it attracts organic traffic without seeding, attributes it correctly, and records it through the full pipeline, with the first probes arriving within hours of exposure and no active advertisement. Second, the composition of that traffic is consistent with the design's premises. Activity concentrated on the self-hosted surface and on its most commonly exposed services; a large fraction of sources sent a single request and never returned; and a substantial share of distinct sources belonged to recognized internet-wide scanning services. Table~\ref{tab:agg} summarizes the window, and Figure~\ref{fig:daily} shows the daily volume, a slowly rising baseline punctuated by short campaign bursts. We report these as characteristics of the observed traffic, not as estimates of the surface at large.
 
\begin{table}[h]
\centering
\footnotesize
\caption{Deployment window data statistics}
\label{tab:agg}
\renewcommand{\arraystretch}{1.2}
\begin{tabular}{@{}lr@{}}
\toprule
\textbf{Quantity} & \textbf{Value} \\
\midrule
Observation window & \obswindow \\
Total requests recorded & \nreq \\
Distinct source addresses & \nsrc \\
Source countries / source ASNs & \ncountry / \nasn \\
Single-request sources & \pctsingle \\
Two most common request types (model list, root) & \pcttoptwo of requests \\
Top-5 ASNs, share of requests & \pcttopfiveasn \\
Requests carrying coded payloads (LiteLLM facade) & \ncoded \\
Distinct payload builds recovered & \nbuilds \\
Honeytoken authentications / canary echoes & 0 / 0 \\
\bottomrule
\end{tabular}
\end{table}
 
\noindent\footnotesize{ASNs = distinct operator-controlled IP blocks.}\normalsize
 
\subsection{Model Enumeration}
 
\begin{figure}[h]
\centering
\includegraphics[width=1.\linewidth]{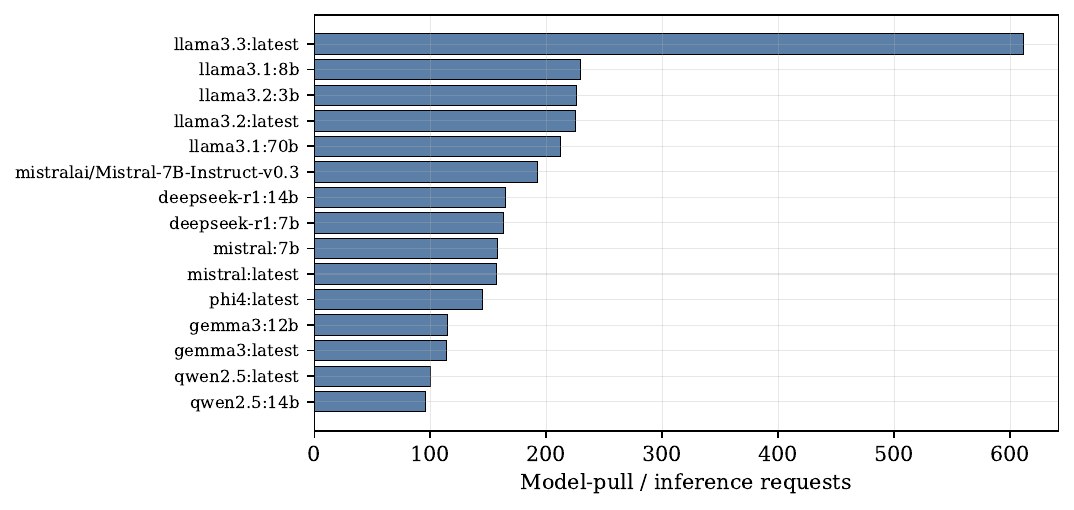}
\caption{Distribution of models requested at the Ollama facade during the study window}
\Description{Horizontal bar chart showing model-pull and inference
requests observed at the Ollama facade. Llama 3.x variants account for
the largest shares, followed by requests for Mistral, DeepSeek-R1,
Phi, Gemma, and Qwen variants.}
\label{fig:models}
\end{figure}

The deployment surfaces a result of independent interest that is a direct consequence of \principle{}. Requests that pull or name a model at the Ollama facade expose the specific inventory an adversary is cataloging, and because the facade answers them faithfully without loading anything, it captures the full distribution of requested models (Figure~\ref{fig:models}). That distribution is dominated by current open-weight families, with Llama 3.x, Mistral, DeepSeek-R1, and Qwen variants accounting for the bulk of requests. The concentration on models small enough to run on consumer GPU hardware suggests that the actors probing exposed model servers are enumerating capacity for cheap, widely deployable inference rather than targeting frontier-model capability specifically. The presence of DeepSeek-R1 and Qwen variants, which are among the most widely deployed open-weight families globally, is consistent with broad-scope tooling targeting available capacity rather than geographically selective targeting. One selection effect qualifies this: an Ollama facade attracts tooling that already expects small, self-hostable models, so the concentration on consumer-GPU models is partly an artifact of the decoy's identity, not a property of the adversary population. We offer this as an illustrative early finding rather than a quantified claim: faithful emulation of the model-pull protocol yields the target list as a byproduct, with no model ever present.
 
\subsection{Captured Credential-Harvesting Payload}
\label{sec:payload}
 
\begin{figure*}[t]
\centering
\begin{tikzpicture}[
  font=\footnotesize,
  life/.style={draw=gray!40, very thin},
  actual/.style={-{Stealth[length=2mm,width=1.4mm]},
                 line width=0.8pt, draw=gray!75!black},
  blocked/.style={-{Stealth[length=2mm,width=1.4mm]},
                  line width=0.7pt, draw=gray!40, dashed},
  step/.style={circle, fill=#1, text=white,
               font=\bfseries\scriptsize,
               minimum size=4.5mm, inner sep=0pt},
  block/.style={draw=#1!60, fill=#1!8,
                rounded corners=2pt,
                inner sep=2pt,
                font=\scriptsize,
                text width=29mm,
                align=center},
  inferd/.style={draw=orange!60, fill=orange!6,
                 rounded corners=3pt,
                 font=\scriptsize,
                 text=orange!80!black,
                 text width=36mm,
                 align=center,
                 inner sep=3pt},
  attackbox/.style={draw=gray!40, fill=gray!4,
                    dashed,
                    rounded corners=2pt,
                    font=\scriptsize,
                    text width=37mm,
                    align=left,
                    inner sep=3pt}
]

\node[block=gray] (attacker) at (0,0) {Attacker};

\node[block=blue] (gateway) at (6,0)
  {\textbf{LiteLLM gateway}\\
   \textit{\scriptsize (InferD emulator)}};

\node[block=gray] (host) at (12,0)
  {Host OS + Cloud};

\draw[life] (0,-0.5) -- (0,-7.2);
\draw[life] (6,-0.5) -- (6,-7.2);
\draw[life] (12,-0.5) -- (12,-7.2);

\draw[actual]
  (0,-1.0) --
  node[step=gray!55!black, pos=.55, above=3pt]{1}
  (6,-1.0);

\node[
  font=\scriptsize,
  align=left,
  text width=48mm,
  anchor=north
] at (3,-1.15)
{
  \texttt{POST /test/connection}\\
  \texttt{\{transport: stdio, command: python3,}\\
  \texttt{args: [<b64\_blob>]\}}
};

\node[inferd] at (6,-2.30)
{
  \textbf{InferD intercepts}\\
  flags \texttt{base64\_payload} and
  \texttt{command\_field};\\
  payload logged verbatim
};

\draw[blocked]
  (6,-3.25) --
  node[step=gray!45, pos=.82, below=-3pt]{2}
  (12,-3.25);

\node[
  font=\scriptsize\itshape,
  text=gray!65,
  anchor=south
] at (9.8,-3.17)
  {blocked: static-pool emulator};

\node[
  font=\scriptsize,
  text=gray!65,
  anchor=north
] at (9,-3.35)
  {spawn \texttt{python3 -c <decoded>}};

\node[attackbox] at (12,-4.60)
{
  \textbf{3a} Local: \texttt{/proc/[pid]/environ},
  config, \texttt{.env}, \texttt{/run/secrets}\\[2pt]

  \textbf{3b} Cloud identity:
  IMDS, ECS/EKS, Azure MI\\[2pt]

  \textbf{3c} Secret managers:
  Vault, Key Vault, SSM, GCP SM,
  Doppler, Infisical
};

\draw[blocked]
  (12,-5.85) --
  node[step=gray!45, pos=.15, below=3pt]{4}
  (6,-5.85);

\node[
  font=\scriptsize,
  text=gray!65,
  anchor=south
] at (9,-5.85)
  {credentials inserted into \texttt{description}};

\draw[actual]
  (6,-6.65) --
  node[step=blue!60, pos=.12, above=3pt]{5}
  (0,-6.65);

\node[
  font=\scriptsize,
  align=center,
  text=orange!80!black,
  anchor=south
] at (3,-6.57)
{
  \texttt{HTTP 200 tools/list}\\
  static-pool response; no credentials returned
};

\end{tikzpicture}

\caption{MCP \texttt{stdio} as a code-execution primitive and
\sys{}'s interception. Solid arrows show the observed interaction,
while dashed gray elements show the prevented execution path.
\sys{} intercepts the request and returns a static response; the
process spawn (step~2) and subsequent credential-access stages
(3--4) therefore never execute. Kernel egress filtering provides
an additional backstop for the cloud-access paths (3b, 3c).}
\Description{Sequence diagram involving an attacker, the LiteLLM
gateway emulated by InferD, and the host operating system and cloud
environment. The observed request sends a Python command with a
base64-encoded payload to the test-connection endpoint. InferD detects
and logs the payload and returns a static tools-list response. A dashed
prevented path shows the subprocess spawn, local and cloud credential
access, secret-manager access, and credential insertion that would
have occurred on a vulnerable real service but do not execute in
InferD.}

\label{fig:mcp-attack}
\end{figure*}
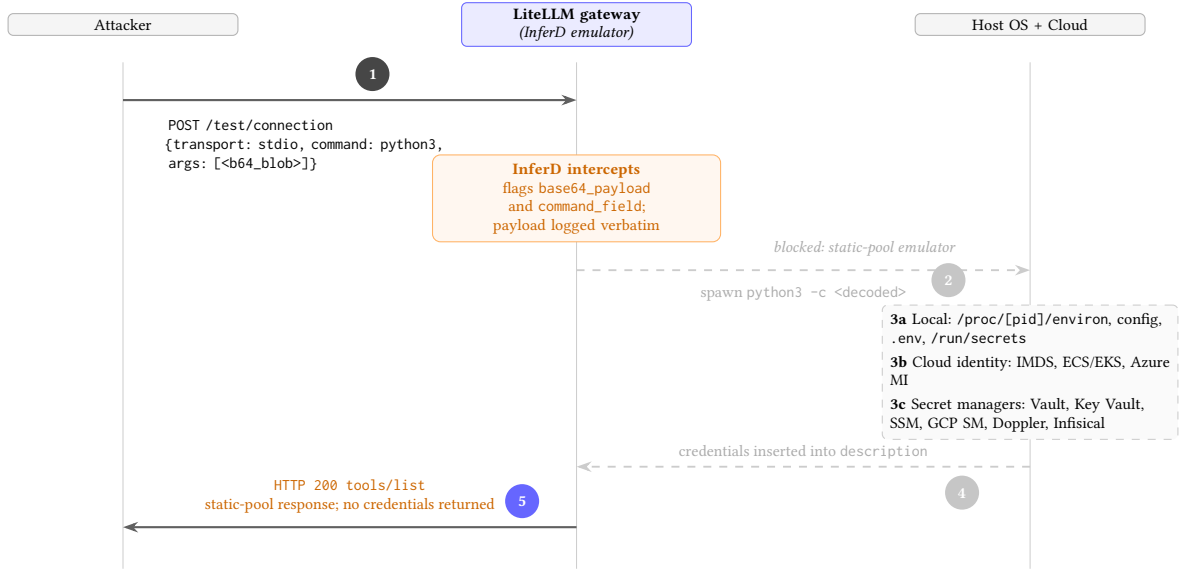

The vulnerability surface layer and the structural anomaly scan (Section~\ref{sec:principle}) jointly produced the most substantive empirical finding of the deployment window. Across the observation period, three probe sources sent a combined \ncoded{} requests to the LiteLLM-emulated facade, distributed across \nlitepath{} distinct endpoint-path variants, including health-check, versioned-API, and double-slash path forms, and two request actions: \texttt{test/connection} and \texttt{test/tools/list}. Each request carried an embedded, base64-encoded Python payload delivered via the MCP stdio transport convention introduced above: a JSON body specifying \texttt{transport: stdio}, \texttt{command: python3}, and an \texttt{args} field containing a self-executing base64 blob. The \texttt{test/connection} endpoint is a legitimate LiteLLM feature for verifying that a configured tool backend is reachable; when invoked with the \texttt{stdio} transport, it accepts an arbitrary command specification from the request body and is designed to spawn the specified process to test connectivity. In a real deployment this constitutes a code-execution primitive: any caller with access to the endpoint can supply an arbitrary subprocess command. \sys{} emulates the endpoint response without executing anything (Figure~\ref{fig:mcp-attack}).
 
\paragraph{Payload structure.} Static analysis of the decoded payloads, recovered by both anomaly-flag-driven queries (flags \texttt{base64\_payload}, \texttt{command\_field}, \texttt{code\_in\_field}) and an independent flag-agnostic base64 scan of every stored request body (see Appendix~\ref{app:detect}), revealed a purpose-built cloud-credential harvester structured as an ordered extraction pipeline rather than a single-vulnerability exploit, making it agnostic to the specific mechanism by which code execution was obtained. The pipeline proceeds through eight stages: environment enumeration across all process \texttt{/proc/[pid]/environ} entries on the host, prioritizing processes whose command line references LiteLLM and checking approximately 80 hardcoded high-value variable names spanning providers including OpenAI, Anthropic, AWS, Azure, Slack, Stripe, and HashiCorp Vault; configuration-file parsing extracting the \texttt{--config} argument from the LiteLLM process and reading the referenced YAML; mounted-secret path traversal over \texttt{/run/secrets}, \texttt{/var/run/secrets}, and \texttt{/app/secrets}; value-pattern detection by credential string prefix independent of variable naming, covering approximately two dozen formats such as \texttt{sk-proj-}, \texttt{sk-ant-api03-}, \texttt{xai-}, \texttt{AIzaSy}, and \texttt{hvs.}; ambient cloud-identity token acquisition via AWS EC2 IMDS (both IMDSv2 and v1 fallback), ECS/Fargate task metadata, EKS pod-identity, and in the later variant Azure Instance Metadata Service; secret-manager traversal using the acquired identity across HashiCorp Vault, Azure Key Vault, GCP Secret Manager, AWS SSM Parameter Store, Doppler, and Infisical; Kubernetes API enumeration via the pod's mounted service-account token (later variant only); and exfiltration via the protocol response channel described below. One implementation detail worth noting for pattern-matching purposes is that the cloud-metadata address 169.254.169.254 is constructed at runtime via \texttt{struct.pack(">I", 0xA9FEA9FE)} rather than appearing as a literal string in the source, consistent with evading naive string-based static detection, though its effect against this deployment's payload-inspection tooling was immaterial.
 
\paragraph{Exfiltration channel.} Rather than opening an outbound connection to a separate collection endpoint, the harvested data is serialized into the \texttt{description} field of a fabricated MCP \texttt{tools/list} response and returned inline to the caller. The attack is entirely in-band within the MCP protocol: no separate command-and-control channel is opened, and network-egress monitoring that relies on identifying anomalous outbound connections would not observe this exfiltration path. Under \sys{} this distinction is immaterial: the emulator answered all \ncoded{} requests from the static pool and never spawned the process, so no stage ran, and the egress rule would in any case have blocked any outbound leg. The exfiltration mechanism is recorded as intent, which is precisely what the non-execution principle guarantees. The design implication for detection of the real services \sys{} emulates is that credential exfiltration via in-band protocol channels requires monitoring of response content rather than network egress, a point we return to in Section~\ref{sec:discuss}.
 
\paragraph{Version drift.} Two distinct payload builds were identified by SHA-256 of decoded content (Table~\ref{tab:payload}), related as a strict superset with a sequence similarity of 0.929, confirming common lineage rather than independent construction. Variant~A was observed from a single source in AS37693 (Ooredoo Tunisie, mobile carrier). Variant~B, which adds the Azure Instance Metadata Service and Kubernetes API enumeration modules, appeared three days later from two sources in AS14061 (DigitalOcean, Germany). All three sources shared an identical HTTP User-Agent string, a non-generic string consistent with a scanning tool rather than a standard browser or library default, despite spanning two structurally different network types; this is a stronger indicator of shared tooling than IP co-location alone, but does not establish that a single operator was responsible for both sessions. None of the three sources matched the published ASNs of known internet-wide research-scanning projects such as Censys (AS398722) or Shodan (AS212238), and none carried a PTR record.

\begin{table}[h]
\centering
\footnotesize
\caption{Characteristics of the two credential-harvester payload builds captured from the LiteLLM facade. Sequence similarity is computed over decoded source content.}
\label{tab:payload}
\renewcommand{\arraystretch}{1.2}
\begin{tabular}{@{}lll@{}}
\toprule
\textbf{Characteristic} & \textbf{Variant A} & \textbf{Variant B} \\
\midrule
SHA-256 prefix (decoded) & \texttt{ef111d4b} & \texttt{c1dfa178} \\
Decoded size (bytes) & 57{,}414 & 66{,}176 \\
Sequence similarity to Variant A & baseline & 0.929 \\
\midrule
Azure IMDS module & absent & present \\
Kubernetes API enumeration & absent & present \\
\midrule
Source ASN & AS37693 (Tunisia) & AS14061 (Germany) \\
Temporal relation & observed first & observed 3 days later \\
\bottomrule
\end{tabular}
\end{table}
 
\paragraph{Scientific position.} Several limitations bound what can be inferred from this observation. The sample is small: three sources, a eight-week window, and a single vantage point preclude any population-level claim. Attribution of the two network-type groups to a single operator is not established; shared tooling and independent reuse are equally consistent with the evidence. The infrastructure metadata is suggestive of operational rather than research scanning but does not constitute attribution. What the observation does establish concretely is that a purpose-built, cloud-provider-aware post-exploitation tool is being delivered to exposed LiteLLM endpoints using an MCP-specific code-execution primitive, that the tool is under active development within a detectable timeframe, and that inferD's structural anomaly scan correctly flagged all \ncoded{} instances at capture time without any signature specific to this payload. Cross-referencing all three source addresses against every honeytoken and canary record in the dataset returned zero matches, consistent with the non-execution guarantee: no credential, real or planted, existed for the attacker's pipeline to harvest and subsequently reuse.

\subsection{Conditional Planes and Their Activation Requirements}
The provenance and agentic planes, which capture the signals the framework most wants, produced nothing in the window: no honeytoken authentications, no canary echoes, and no populated multi-step tool-call sessions. These planes are conditional instruments: the provenance plane can only register credential abuse once planted credentials are in circulation to be stolen, and the agentic plane can only populate when agentic traffic reaches the decoy, which over a short single-vantage window is rare. The null also has a design-level cause: A5 actors are precisely those able to run the reasoning probe of Section~\ref{sec:detect} and disengage on detecting a static-pool facade, so these planes bound sophisticated activity from below rather than measure it, even under longer collection. The null is therefore primarily a statement about the deployment, not about the world, and it identifies the two concrete steps, active honeytoken seeding and longer multi-vantage collection, that a measurement study built on this instrument would have to take. Reporting these null results rather than eliding them is part of treating the deployment as a pipeline validation rather than a measurement study: the instrument confirms end-to-end function and makes explicit what the conditional planes need to register signal, which is the prerequisite for a follow-on study.

\section{Discussion and Related Work}
\label{sec:discussion}
 
\subsection{Detectability Limits}
\label{sec:detect}
 
A honeypot design earns trust by confronting how it can be unmasked. The most fundamental tell in \sys{} is the one the principle chooses to accept: the facade does not reason, so an adversary who submits a prompt with a verifiable answer and checks the reply can detect the absence of genuine inference. This is the deliberate cost of non-execution, and it binds only adversaries sophisticated enough to test for reasoning. A second tell is response repetition, which variant expansion mitigates but a patient adversary collecting many responses could still use to estimate pool size \cite{srinivasa2023gotta}. A third is timing: a facade that answers a completion in milliseconds is immediately distinguishable from a real model that requires seconds per generation, the LLM-specific instance of a timing-based detectability category established for classical honeypots \citep{holz2005detecting} and now formalized as a measurable metric for LLM-powered honeypots specifically \citep{vero2026honeyval}, and modeling plausible inter-token latency would close this gap as a natural refinement. A fourth is the transport fingerprint, which may diverge from the genuine provider, including differences in TLS certificate organization fields and HTTP header ordering, and which we log so that any such divergence is visible in our own data as well \cite{vetterl2018bitter}. The decisive tell for the vulnerability and agentic planes is that an adversary with out-of-band confirmation of effect will observe that the effect never occurs. We restate this rather than minimize it: non-execution is, by construction, detectable to anyone who looks for its absence. The design's position is that it is better to observe intent safely than to demonstrate effect dangerously, and the detectability of the safety property is the visible signature of that choice.
 
\subsection{Scope, Extensibility, and Measurement}
\label{sec:discuss}
 
The deployment's limits point directly at the work the instrument is built to enable. Its most valuable planes are conditional, and exercising them defines a measurement agenda rather than a defect: seeding honeytokens where adversaries collect leaked credentials, so that Surface~2 and the provenance plane can register abuse; extending collection across time and across geographically distributed vantage points, so that rarer agentic behavior can accumulate and so that globally uniform scanning can be separated from locality-specific activity; and modeling response latency to narrow one detectability gap. There are limits the design does not address at all: it observes inbound threat traffic only, and says nothing about supply-chain compromise of models or packages, or about activity confined to a victim's internal network.
 
The instrument is designed to extend in three directions as the threat landscape evolves. First, multi-vantage deployment: distributing \sys{} instances across geographically diverse ASNs and coordinating their pseudonymized logs into a shared event stream would allow globally uniform scanning to be separated from locality-specific campaigns, following the distributed observatory model of network telescopes \citep{bailey2006dark,holoscope2025}. Second, surface extension: the pool-and-emulator architecture accommodates new services incrementally; each new CVE or emerging deployment pattern can be absorbed by adding one router and one response pool without modifying shared middleware. Third, active honeytoken seeding: planting \sys{}-specific credential patterns in controlled environments where harvesting tools are known to operate, then observing whether those tokens appear against Surface~2, would activate the provenance plane without requiring passive waiting.
 
\subsection{Related Work}
\label{sec:related}
With the design in view, its relationship to prior work is easier to state. In fact, classical honeypot frameworks established the vocabulary we inherit: that one host can present many convincing services, and that resisting fingerprinting is as important as the emulation itself \citep{provos2004honeyd,spitzner2003honeypots,mokube2007honeypots}, a lineage surveyed at the software level \citep{nawrocki2016survey} and, more broadly, across the full space of honeypots, honeytokens, and moving-target defense \citep{han2018deception,javadpour2024survey}, an idea operationalized for shells and network services by interaction-focused systems such as Cowrie\footnote{Michel Oosterhof, \textit{Cowrie SSH/Telnet Honeypot}, 2017. \url{https://github.com/cowrie/cowrie}}. \sys{} adopts their discipline and their honesty about detectability, but targets a surface, AI inference infrastructure, that they were not built to imitate, and where fidelity is defined against the behavior of a model. Multi-service honeypot platforms such as T-Pot\footnote{Telekom Security, T-Pot: The All-In-One Multi-Honeypot Platform. \url{https://github.com/telekom-security/tpotce}} demonstrate the value of aggregating heterogeneous emulators under a single deployment unit, an architectural parallel to \sys{}'s two-surface model, though none of their constituent emulators was designed for AI-specific protocol fidelity or the non-execution guarantee that AI inference emulation requires.
 
A recent line of work places a language model inside the honeypot as its response generator, obtaining unbounded interaction depth for shells \citep{sladic2024llm,otal2024llm}. \sys{} deliberately inverts this relationship: the model is our subject, not our tool, and Section~\ref{sec:principle} argues why generating responses with a model is the wrong choice for an inference honeypot specifically. Closest in intent is work that instruments generic honeypots to detect autonomous agents through prompt injection and timing analysis \citep{volkov2025llmhoneypot}, from which we adapt the canary technique; that work studies agents \emph{attacking} conventional infrastructure, whereas \sys{} studies attacks \emph{directed at} AI infrastructure. At the industry scale, telemetry has documented mass reconnaissance of exposed model servers\footnote{GreyNoise Intelligence, \textit{Threat Actors Actively Targeting LLMs: Mass Scanning of Exposed AI Infrastructure}, 2026. \url{https://www.greynoise.io/blog/threat-actors-actively-targeting-llms}}, and concurrent work reports a honeypot of comparable two-surface scope whose measurement results are not yet public \footnote{Adam Toscher, \textit{Building Honey-AI, Part 1} (2026). \url{https://www.cobalt.io/blog/building-honey-ai-part-1-why-ai-infrastructure-needs-its-own-honeypot}}; a systematization of honeypots and language models situates these efforts and marks the AI infrastructure surface, treated as a target, as the least developed among them \citep{bridges2025sok}. What distinguishes \sys{} within this landscape is not scope but method: non-execution by design paired with a kernel-enforced egress guarantee, a threat model and emulated surface grounded in the attributable incident record (CVE-2023-48022, CVE-2024-37032, CVE-2024-6587, documented above), and coverage of the agentic surface stated against an external taxonomy.
 
Finally, the phenomena \sys{} is built to observe are documented, so far, mostly under controlled conditions: prompt injection and its indirect variants \citep{greshake2023indirect,liu2023prompt,mchugh2025hybrid}, retrieval poisoning \citep{chen2024agentpoison}, multi-agent propagation \citep{lee2024promptinfection}, and tool poisoning against real MCP servers, which a recent benchmark measures at substantial success rates with little pre-execution protection from alignment \citep{wang2025mcptox,beurer2025design}. \sys{} is designed to observe the operational precursors of these attacks passively, in real traffic, without reproducing them.
 
\section{Conclusion} 
\label{sec:conclusion}
 
AI inference infrastructure has become a distinct, exposed attack surface, and emulating it for observation poses a deception problem conventional honeypots do not face: the facade must reproduce everything an adversary can observe while executing nothing an adversary submits. \sys{} resolves this with a single principle, \principle{}, and we have argued that the principle is the right choice rather than a concession, because it avoids reintroducing the model whose abuse the honeypot studies, and because it lets non-execution by design be backed by an egress guarantee the kernel enforces regardless of emulator correctness. Around that principle the framework composes four deception planes and a provenance plane, each developed with its rejected alternative and its detectability limit in view, and realized with the systems techniques that make it faithful and safe on commodity hardware. An initial deployment confirms the instrument functions end to end, surfaces a first empirical finding in the form of a novel credential-harvesting payload delivered via an MCP stdio code-execution primitive.


\begin{acks}

This work was conducted as part of the NGSOTI project, co-funded by
the European Union's Digital Europe Programme under Grant Agreement
No.~101127921 (DIGITAL-ECCC-2022-CYBER-03).

\end{acks}

\section*{Availability}
The source code and supporting artifacts for \sys{}, including the emulator code, response-pool schema, and event schema, are publicly available at: \url{https://github.com/hichamhammouchi/InferD}.

\section*{Generative AI Disclosure}
AI tools were used for language editing for readability and for reviewing and debugging code. All research design, system implementation, and analysis artifacts are authors' own.
 
\bibliographystyle{ACM-Reference-Format}
\bibliography{references}
 
\appendix
 
\section{Payload Detection Methodology}
\label{app:detect}
 
Two complementary methods were applied to recover coded payloads from the stored request dataset. The primary method queries events flagged at capture time by the structural anomaly scan, which assigns flags including \texttt{code\_in\_field}, \texttt{shell\_metachar}, \texttt{command\_field}, \texttt{pickle\_payload}, and \texttt{base64\_payload} based on lexical inspection of each request's body and path fields at ingest. The secondary method applies an independent, flag-agnostic scan of every stored request body for base64-decodable content, attempting base64 decoding of each 32-byte-minimum substring and applying gzip, bz2, lzma, and zlib decompression to the result; payloads encoded with nested base64 are recovered by a second pass of the same procedure. The two methods agreed on the credential-harvesting payloads documented in Section~\ref{sec:payload} and jointly confirmed no additional undetected code payloads elsewhere in the \nreq{} requests across the \nsvc{} emulated services at the time of analysis. Three additional base64-decodable blobs surfaced in the flag-agnostic scan on an unrelated service and decoded to PNG image data submitted as multimodal chat content, confirmed benign by valid PNG structure verification with no data appended after the terminating chunk. A residual blind spot exists for payloads encoded with schemes outside base64 and the four compressed formats listed above. One capture-pipeline gap was identified separately: probes targeting the Jupyter kernel's code-execution endpoint trigger the same anomaly flags, but the submitted code is not currently persisted in the stored request body for that service and could not be analyzed; persisting this content is a straightforward fix planned for the next deployment iteration.
 
\end{document}